\documentclass{PoS}
\usepackage{wrapfig}
\title{N(N)LO event files: applications and prospects}

\ShortTitle{N(N)LO event files: applications and prospects}

\author{\speaker{Daniel Ma\^{\i}tre}\\
        IPPP Durham\\
        E-mail: \email{daniel.maitre@durham.ac.uk}}

\author{Gudrun Heinrich\\
        Max Planck Institute for Physics\\
       F\"ohringer Ring 6\\
       80805 Munich, Germany\\
        E-mail: \email{gudrun@mpp.mpg.de}}

\author{Mark Johnson\\
        Manchester University and the Cockcroft Institute\\
        E-mail: \email{mark.andrew.johnson@cern.ch}}

\abstract{In this contribution we comment on the use of NLO n-Tuples and consider their extension to NNLO. As an application of n-Tuples we present preliminary results of a strong coupling extraction from a measurement of the production of  $Z+2,3,4$ jets  at the LHC at $7\,\rm TeV$.}

\FullConference{Loops and Legs in Quantum Field Theory\\
		24-29 April 2016\\
		Leipzig, Germany}

\begin{document}

\section{n-Tuples}
The calculation of high multiplicity processes at NLO in now possible for a large number of processes in a largely automated way (see e.g.~\cite{Badger:2016bpw} for a review). While they can be automated, high multiplicity NLO calculations still require a large amount of CPU time to be performed with an acceptable statistical precision. The most expensive part of the calculation is the computation of the matrix elements. Other parts of the calculation such as the evaluation of the parton distribution functions, jet clustering and the construction of observables and differential cross sections to be histogrammed  are typically much less demanding. In practice this also requires to evaluate very similar expressions, differing only by a factor in the factorisation or renormalisation scales or by the PDF used. To avoid  recalculations of the same matrix element many times one can store the matrix elements, the kinematic information and the coefficients of  logarithms involving the renormalisation and factorisation scales in a file and read them back in,  instead of recalculating
the whole matrix element. This strategy, using ROOT \cite{ROOT} as the storage back-end, has been described in \cite{Bern:2013zja}. 

Besides the computational efficiency, this strategy has the advantage of facilitating the dissemination of the NLO results. The clear disadvantage is that the event files tend to be very large. The  advantages outweigh this disadvantage for NLO calculations and this strategy has been used to produce NLO predictions \cite{Z4,W4,pureQCD,Greiner:2015jha,Badger:2013yda, Badger:2013ava} for many Standard Model measurements \cite{WD0,W7TeVCMS,Z7TeVCMS,Jets8TeVAtlas,RratioAtlas,W7TeVAtlas,Z7TeVAtlas}.   

It is natural to wonder whether a similar approach would facilitate the calculation and dissemination of NNLO calculations. The advantages and disadvantages are the same: the matrix elements are computationally expensive to calculate and would benefit from being stored but the amount of events needed would lead to large file sizes. A first study in ref.~\cite{Badger:2016bpw} investigated the size of NNLO event files using the {\tt EERAD3} program \cite{EErad3}.

\begin{figure}
  \begin{center}
    \includegraphics[scale=0.4]{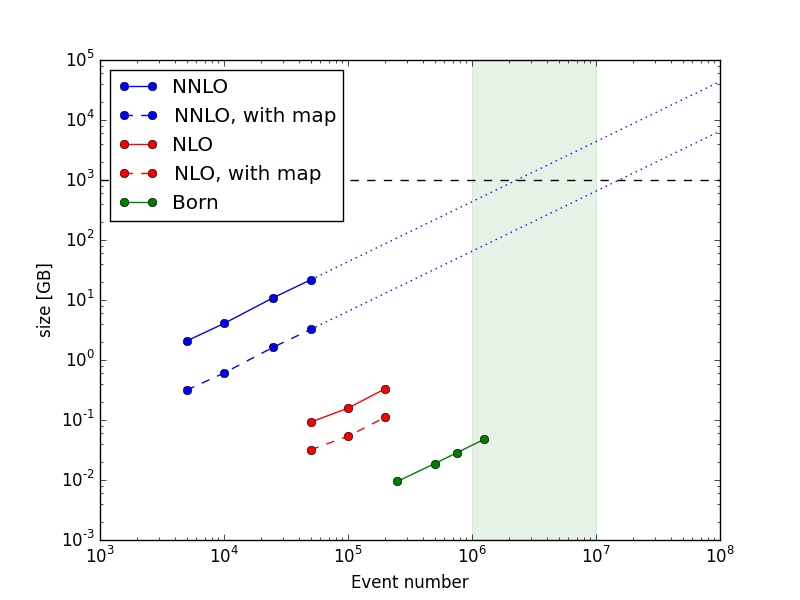}
  \end{center}
\caption{Event file size for $e^+e^-\rightarrow 3\,\rm{jets}$ as a function of the number of events. The green band represents the number of points needed for a reasonable statistical error. The horizontal dashed line is the $1\,\rm{TB}$ threshold which corresponds to the order of magnitude of storage for NLO processes.}\label{fig:sizes}
\end{figure}

One significant change encountered when going from NLO to NNLO is the increase in the number of subtraction terms, which leads to a large number of phase-space configurations that need to be stored in the event file. Instead of storing each set of mapped momenta, one can store the original momenta they are mapped from, along with a reference to the procedure with which the mapped momenta are arrived at. This saves about an order of magnitude of storage space, at the cost of delegating the calculation of the mapped momenta to the user of the event file. Figure~\ref{fig:sizes} shows the storage sizes as a function of the number of phase-space points. 

Based on the extrapolation in figure~\ref{fig:sizes} the usage of NNLO event files appears feasible in this case. The study in \cite{Badger:2016bpw} used a leptonic initial state. One can anticipate that processes with hadronic initial states will be more challenging and require higher statistics, as there are additional integrations over the initial momentum fraction of the partons. This can make the prospects of using event files less favourable. However, there are several reasons to think hadronic processes might still be tractable: a) so far NNLO calculations have been optimised for CPU time, one could optimise them with respect to storage space (using the mapping information is an example of such an optimisation), b) new subtraction techniques such as $q_T$-subtraction \cite{Catani:2007vq} and $n$-jettiness \cite{Stewart:2010tn,Gaunt:2015pea,Boughezal:2015dva} offer an easier subtraction structure which is closer to that of two separate NLO calculations than ``genuine''  NNLO subtraction schemes \cite{GehrmannDeRidder:2005cm,Currie:2013vh,Czakon:2010td}. This ``NLO-like'' structure should lead to more efficient storage possibilities.  

\section{Strong coupling determination}

\begin{figure}
  \begin{center}
    \includegraphics[scale=0.45]{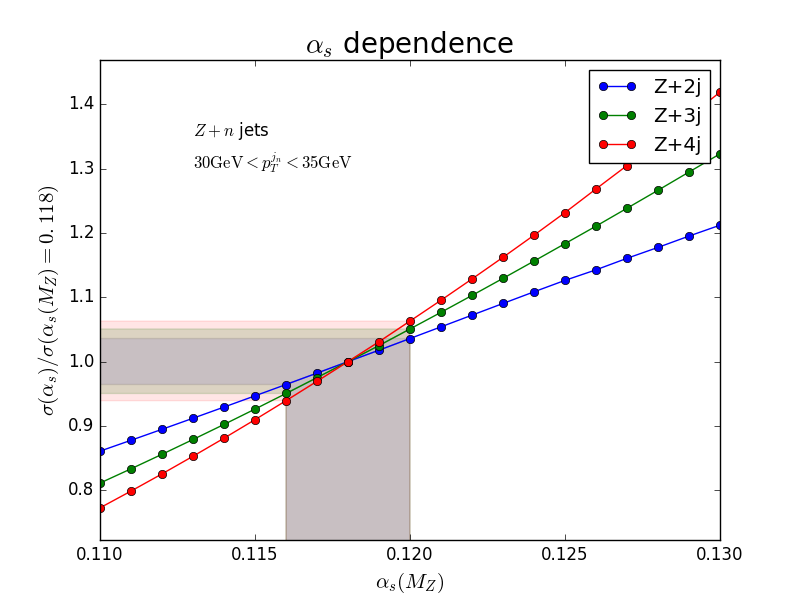}
    \end{center}
  \caption{Left-hand panel: $\alpha_S(M_Z)$ dependence of the value of the $30\,{\rm GeV}< p_T < 35\,{\rm GeV}$ bin in the transverse momentum distribution for the $n$-th jet for $n=2,3,4$.This figure illustrates the fact that a larger uncertainty on the cross section for a higher-multiplicity process can result in the same uncertainty on the predicted values of $\alpha_S(M_Z)$. }\label{fig:alpha}
\end{figure}

The ability to recalculate the NLO predictions for a high multiplicity process relatively cheaply offers new opportunities for these processes to be used in phenomenological applications. We explore one of them in this section, namely extracting the value of the strong coupling constant from $Z+2,3,4\;\mbox{jets}$ measurements. The reason for using high multiplicity processes is that they display an increasingly strong dependence on $\alpha_S(M_Z)$ as the multiplicity increases, as illustrated in figure~\ref{fig:alpha}.

  \begin{figure}
  \begin{center}
    \includegraphics[scale=0.35]{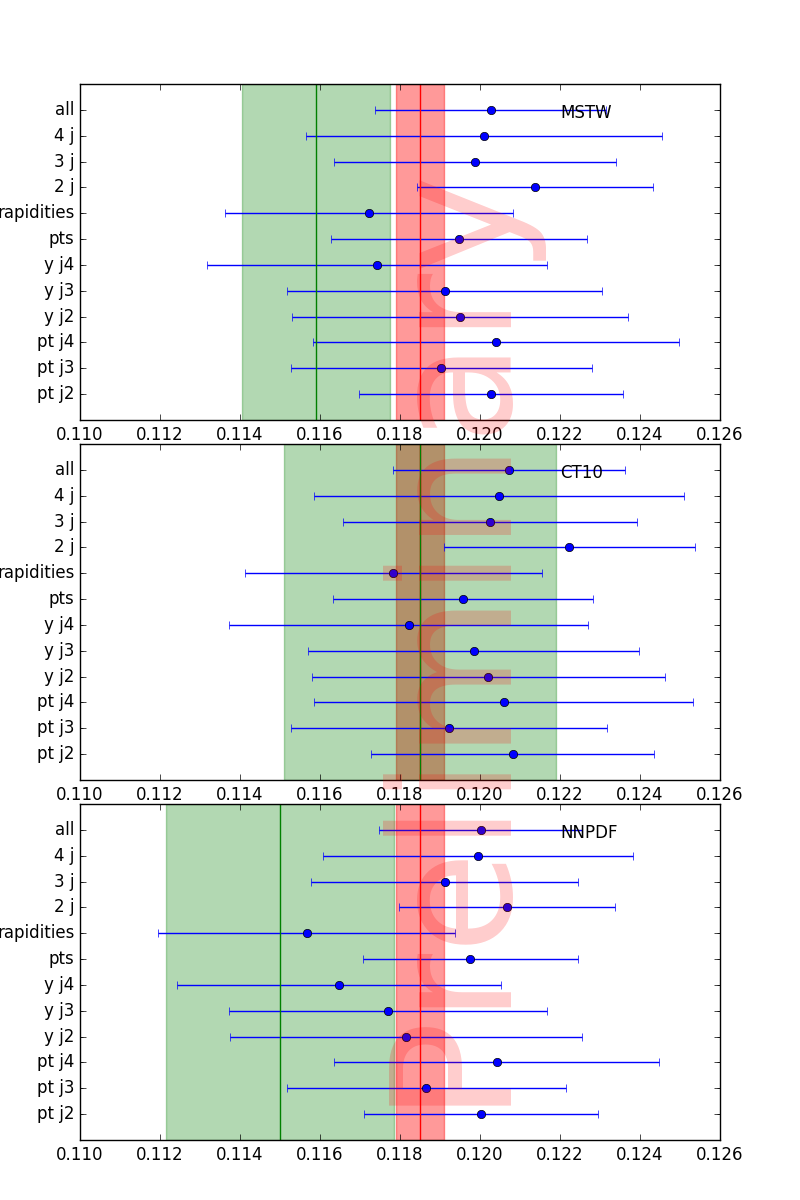}
    \end{center}
  \caption{ Strong coupling values for different PDF sets for each considered histogram and combinations. The green band is the result from \cite{Khachatryan:2014waa}. The red band is the world average~\cite{PDG}. The boundaries of the blue error bars represent the values of $\alpha_s(M_Z)$ for which the $\chi^2-\chi^2_{min}=1$. These error estimates do not include scale variation. }\label{fig:all}
  \end{figure}

For our $\alpha_S$ extraction we use the data from~\cite{Khachatryan:2014waa} for the rapidity and transverse momentum distribution of the $n$-th jet in $Z+n\;\rm jets$ events. We compare it with predictions obtained from the {\tt BlackHat+Sherpa} collaboration~\cite{Z4}. We used the n-Tuples they provide to generate a {\tt fastNLO} grid~\cite{fastNLO} to speed up the calculation of the PDF covariance matrices and scale variations. The fit is obtained by minimising the $\chi^2$ function
\[\chi^2(\alpha_S)=\left(y_{t}(\alpha_s)-y_{d}\right)^TC^{-1}\left(y_{t}(\alpha_s)-y_{d}\right)\;,\]
where $y_t$ is the theory prediction and $y_d$ are the experimental values. The covariance matrix $C$ is given by
\[C=C_{exp}+C_{pdf}+C_{theory}\;,\]
with $C_{exp}$ the experimental error covariance matrix, $C_{pdf}$ the PDF covariance matrix obtained using the LHAPDF library \cite{LHAPDF} and  $C_{theory}$ the statistical covariance matrix of the theory prediction.

Figure~\ref{fig:all} shows the preliminary results for the three PDF sets {\tt MSTW2008}~\cite{MSTW2008}, {\tt CT10}~\cite{CT10} and {\tt NNPDF~2.3}~\cite{NNPDF23}. These sets were chosen to facilitate the comparison with the results obtained in ref.~\cite{Khachatryan:2014waa}. The results are given for fits to different sets of histograms:  a) to the transverse momentum and rapidity of the $n$-th $p_T$-ordered jet for each multiplicity individually, b) for the combination of the transverse momentum and rapidity for each multiplicity, c) for the combination of all three transverse momentum distributions, d) for the combination of all the rapidity distributions, e) for a combination of all histograms.
\section{Conclusion}
In this contribution we reported on an investigation of the suitability of the n-Tuple strategy for NNLO calculations. This strategy seems tractable for processes with no hadronic initial states and while the jury is still out for the case of hadronic initial states we have good reasons to think it will also be tractable in these cases. We also presented preliminary results for an extraction of the strong coupling constant from high multiplicity processes. The accuracy obtained for the value of $\alpha_S(M_Z)$ seems comparable with other determinations at the LHC.    

\bibliography{LL2016}
\bibliographystyle{JHEP}

\end{document}